\documentclass[journal=nalefd,manuscript=letter,layout=twocolumn]{achemso}

\usepackage[version=3]{mhchem} % Formula subscripts using \ce{}

\usepackage{soul,color}

\usepackage{amsmath,amssymb}

\newcommand*\xbar[1]{%
  \hbox{%
    \vbox{%
      \hrule height 0.5pt % The actual bar
      \kern0.5ex%         % Distance between bar and symbol
      \hbox{%
        \kern-0.2em%      % Shortening on the left side
        \ensuremath{#1}%
        \kern-0.0em%      % Shortening on the right side
      }%
    }%
  }%
}

\author{Valeriy~A.~Slipko}
\affiliation{Institute of Physics, Opole University, Opole 45-052, Poland}
\email{vslipko@uni.opole.pl}
\author{Yuriy~V.~Pershin}
\affiliation{Department of Physics and Astronomy, University of South Carolina, Columbia, SC 29208 USA}
\email{pershin@physics.sc.edu}

\title[]{A probabilistic model of resistance jumps in memristive devices}

\abbreviations{IR,NMR,UV}

\begin{document}

\begin{tocentry}
\begin{center}
    \includegraphics[height=4.5cm]{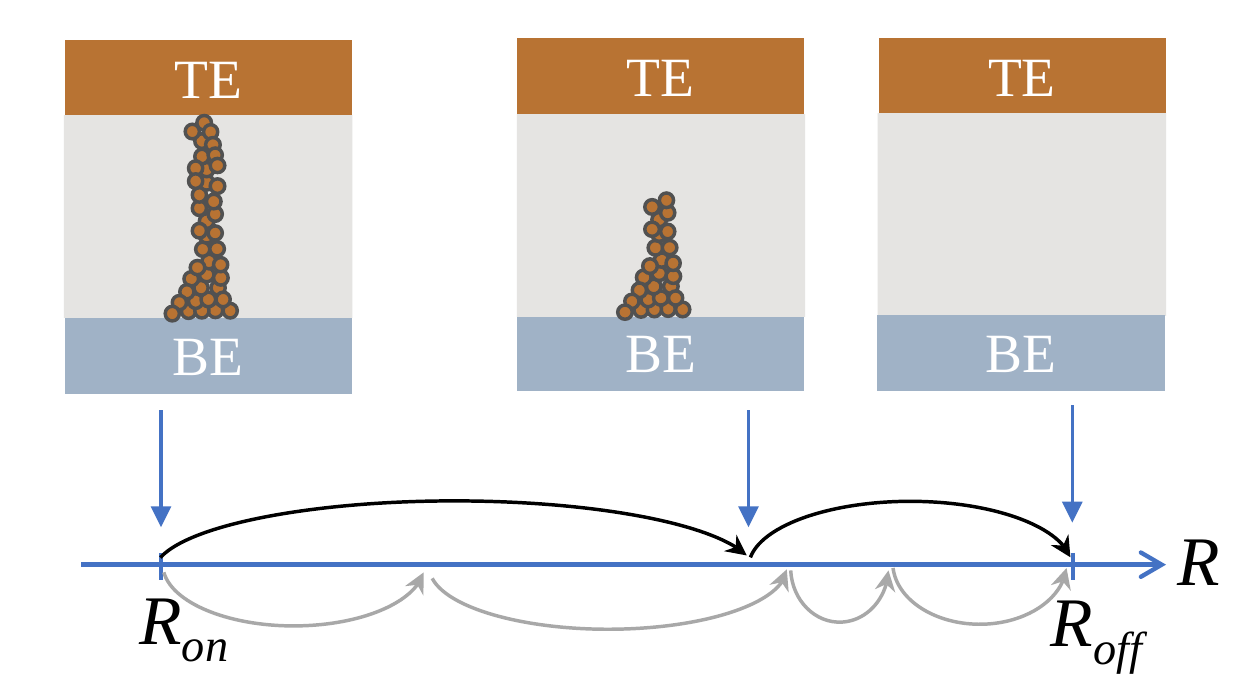}
\end{center}
\end{tocentry}

\begin{abstract}
  Resistance switching memory cells such as electrochemical metallization  cells and valence change mechanism  cells have the potential to revolutionize information processing and storage. However, the creation of {\it deterministic} resistance switching devices is a  challenging problem that is still open. At present, the modeling of resistance switching cells is dominantly based on deterministic models that fail to capture the cycle-to-cycle variability intrinsic to these devices. Herewith we introduce a state probability distribution function and associated  integro-differential equation to describe the switching process consisting of a set of stochastic jumps. Numerical and analytical solutions of the equation have been found in two model cases. This work expands the toolbox of models available for  resistance switching cells and related devices, and enables a rigorous description of intrinsic physical behavior not available in other models.
\end{abstract}


\vspace{1cm}

Resistance switching memory cells (also known as memristive devices~\cite{chua76a}) are emerging components with memory that find applications in neuromorphic~\cite{li2018,xia2019memristive,Christensen_2022}, logic~\cite{vourkas2016emerging,sebastian2020memory}, and reservoir computing circuits~\cite{kulkarni2012memristor}, to name a few. Over the last decade, significant progress has been achieved in the above mentioned and related areas, and includes
the demonstration of high-density crossbar arrays~\cite{pi2019memristor}, diffusive memristors~\cite{wang2017memristors}, etc.
Moreover, the utilization of natural materials and unconventional fabrication techniques~\cite{Kim22a} may lead to memory cells that have a lower carbon footprint compared to silicon-based electronics.

\begin{figure}[tb]
  \includegraphics[width=0.8\columnwidth]{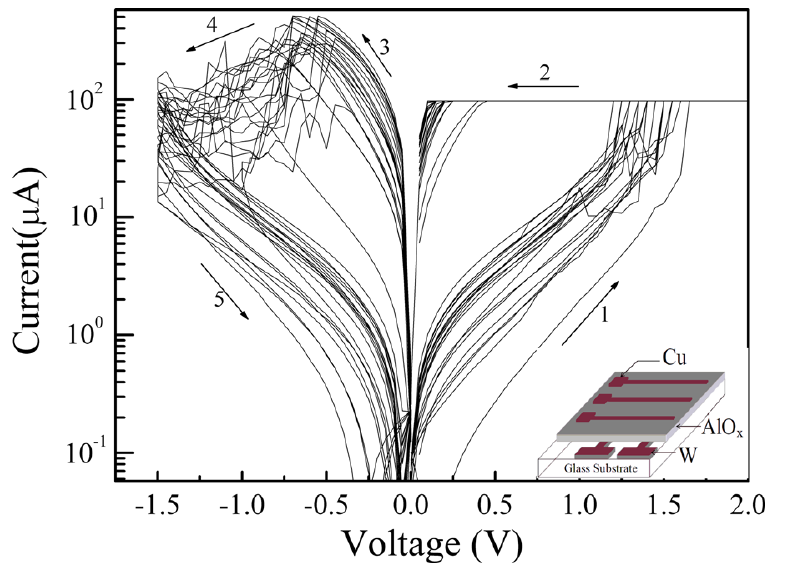}
    \caption{Experimental current-voltage curves of Cu/AlO$_x$/W nonvolatile memory structure (for details, see Ref.~\citenum{Sleiman13a}).  Reprinted from Ref.~\citenum{Sleiman13a}.}
    %, with the permission of AIP Publishing.}
  \label{fig:1}
\end{figure}

Traditionally, memristive devices have been described in terms of deterministic models. The dynamical system approach introduced
by Chua and Kang~\cite{chua76a} provides the mainstream theoretical framework (examples of memristive models can be found in Refs.~\citenum{kvatinsky2012,strachan2013state,pershin2011memory}). However, being relatively simple, deterministic models neglect the
stochastic behavior in particular common to electrochemical metallization (ECM) cells~\cite{valov2011} and valence change mechanism (VCM) cells~\cite{yang2012} -- the most typical memristive devices. As an example, Fig.~\ref{fig:1} shows the current-voltage curve for Cu/AlO$_x$/W nonvolatile memory structure~\cite{Sleiman13a}.

In the present paper, we develop a continuous probabilistic model to describe the evolution of stochastic memristive devices. The main assumption in the model is that the resistance switching occurs via random markovian jumps in the continuous space of internal state variable(s). Previously, we have pioneered  the use of a master equation as a tool for examining the response of stochastic memristive devices with discrete states~\cite{dowling2020probabilistic}, implemented this approach in SPICE~\cite{DowlingSPICE}, and developed a theory of circuits combining discrete stochastic memristive devices with reactive components~\cite{slipko2021theory}. Recently, the master equation was used in the analysis of memristive Ising circuits~\cite{Dowling22a}, which are the  electronic circuit realization of the Ising model. The present paper extends the  discrete master equation approach~\cite{dowling2020probabilistic,DowlingSPICE,Ntinas21a} to a more realistic case of continuous internal states.

To take into account the stochasticity,
we introduce the state probability distribution function $p(x,t)$, where $x\in [a,b]$ is the internal state variable responsible for memory, and $p(x,t)\textnormal{d}x$ is the probability to find the state in the interval from $x$ to $x+\textnormal{d}x$ at the time moment $t$. The state probability distribution function is normalized to 1:
\begin{equation}
    \int\limits_a^b p(x,t)\textnormal{d}x=1.
\end{equation}
 %Note that this description can be straightforwardly extended  to several internal state variables and suitable intervals of their change (other than from 0 to 1).

 The first equation in the model is a statistically averaged Ohm's law
\begin{equation}
\left<I \right>=\left< \frac{V}{R(x,V)} \right>\equiv \xbar{R}^{-1}(V)\cdot V ,
\label{eq:m1}
\end{equation}
where $\left<I \right>$ is the mean current, $V$ is the voltage across the device, $R(x,V)$ is the state- and voltage-depend resistance,  and $\xbar{R}(V)=\left(\int_a^bR^{-1}(x,V)p(x,t) \textnormal{d}x\right)^{-1}$ is the harmonic mean resistance. We emphasize that $\xbar{R}(V)\neq \left<R(x,V) \right>$.

The evolution of $p(x,t)$ is represented by an integro-differential equation
\begin{eqnarray}
    \frac{\partial p(x,t)}{\partial t}=\int\limits_a^b\gamma(x',x,V(x'))p(x',t)\textnormal{d}x'
    \nonumber
   \\
    -p(x,t)\int\limits_a^b\gamma(x,x',V(x
    ))\textnormal{d}x',
    \label{eq:m2}
\end{eqnarray}
which is the second equation in the model. Here,
$\gamma(y,z,V(y))$ is the voltage-dependent transition rate density from internal state $y$ to $z$, and $V(y)$ is the voltage across the device in state $y$.
The first term in the right hand side of (\ref{eq:m2}) describes the transitions to $x$ from all other  states, while the second one -- the reverse transitions.
%We emphasize that in a circuit, the voltage across a memristive device may depend on its state, and this fact is taken into the account by the dependence $V(y)$.

The extension of Eqs.~(\ref{eq:m1}), (\ref{eq:m2}) to several internal state variables is discussed in the Supporting Information (SI).

%Using the above equation it is straightforward to show that the integrated probability is conserved as
%\begin{equation}
%    \frac{\textnormal{d} }{\textnormal{d} t}\int\limits_0^\infty r(R,t)\textnormal{d}R=0.
%\end{equation}

%\subsection*{Transition rates}

\begin{figure*}
  \includegraphics[width=0.32\linewidth]{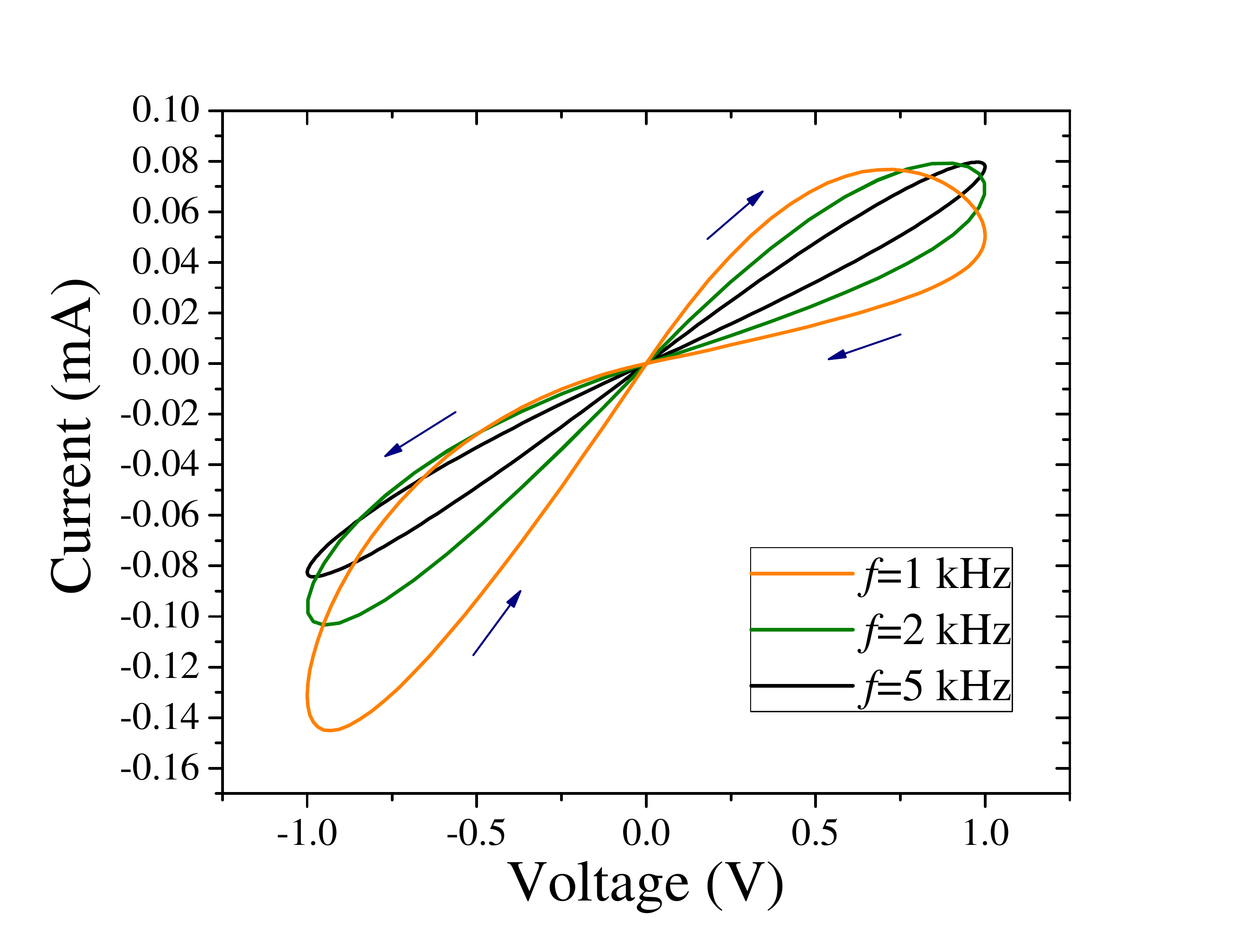}
  \includegraphics[width=0.32\linewidth]{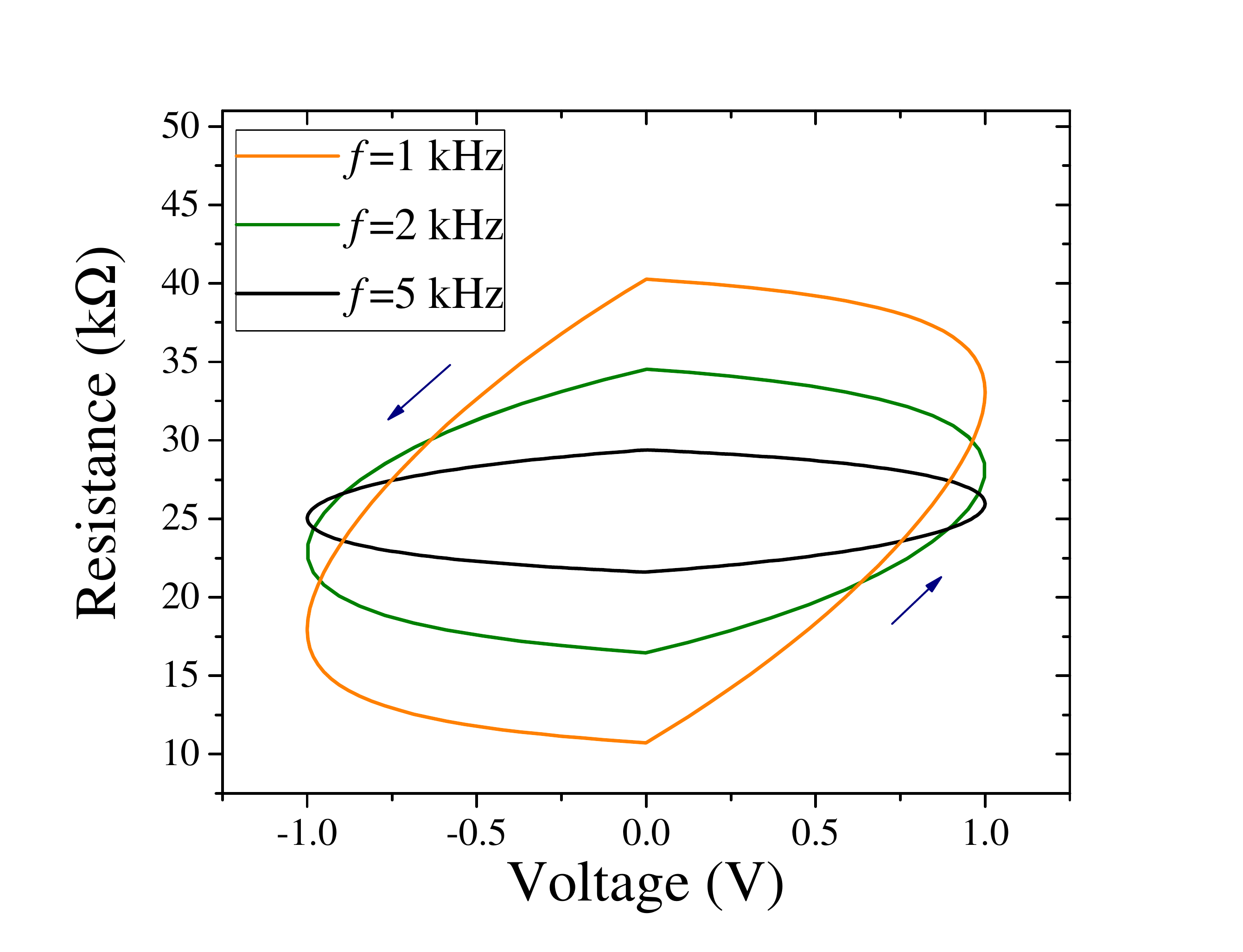}
  \includegraphics[width=0.32\linewidth]{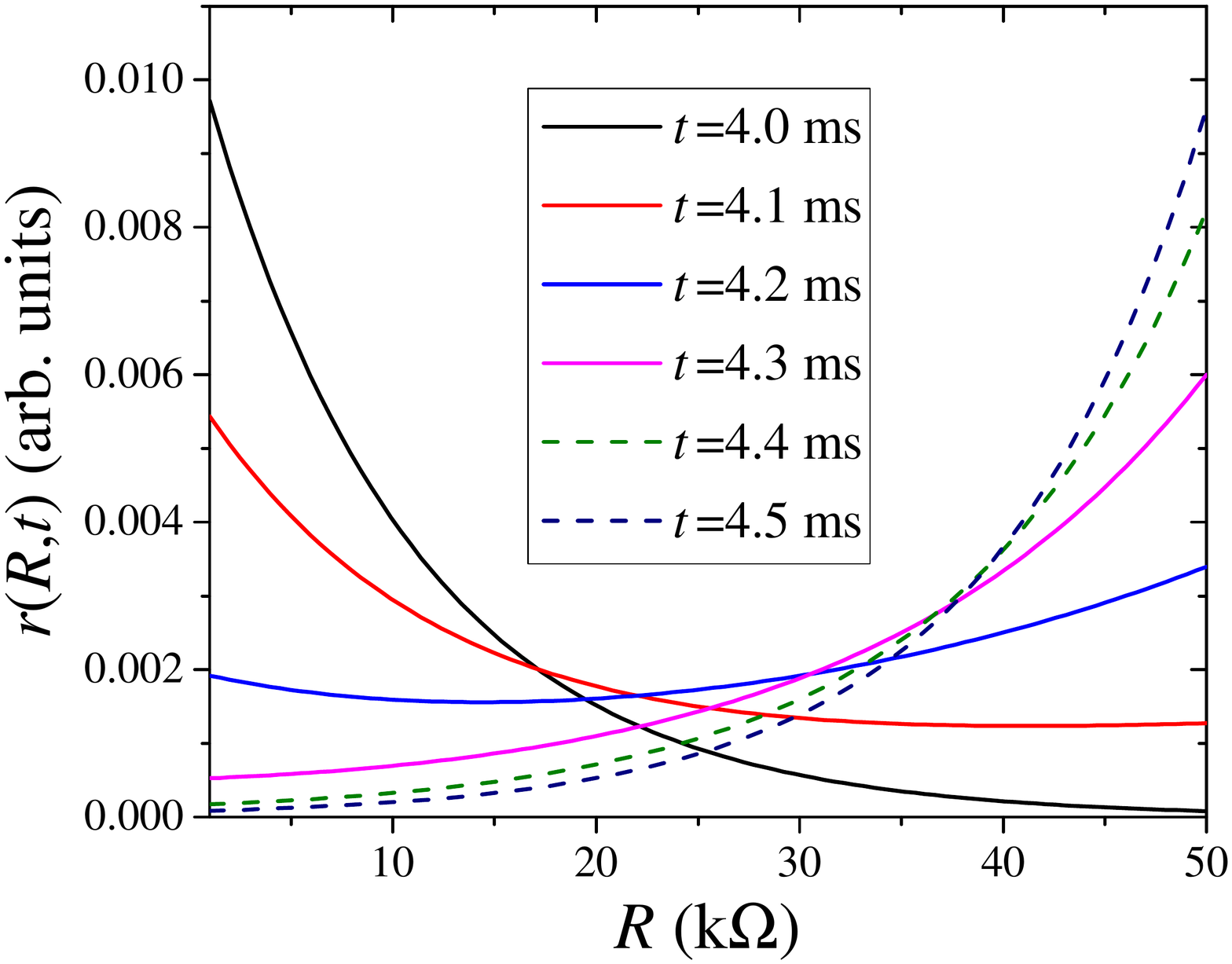}\\
  \centering
  (a) \hspace{5cm} (b) \hspace{5cm} (c)
  \includegraphics[width=0.32\linewidth]{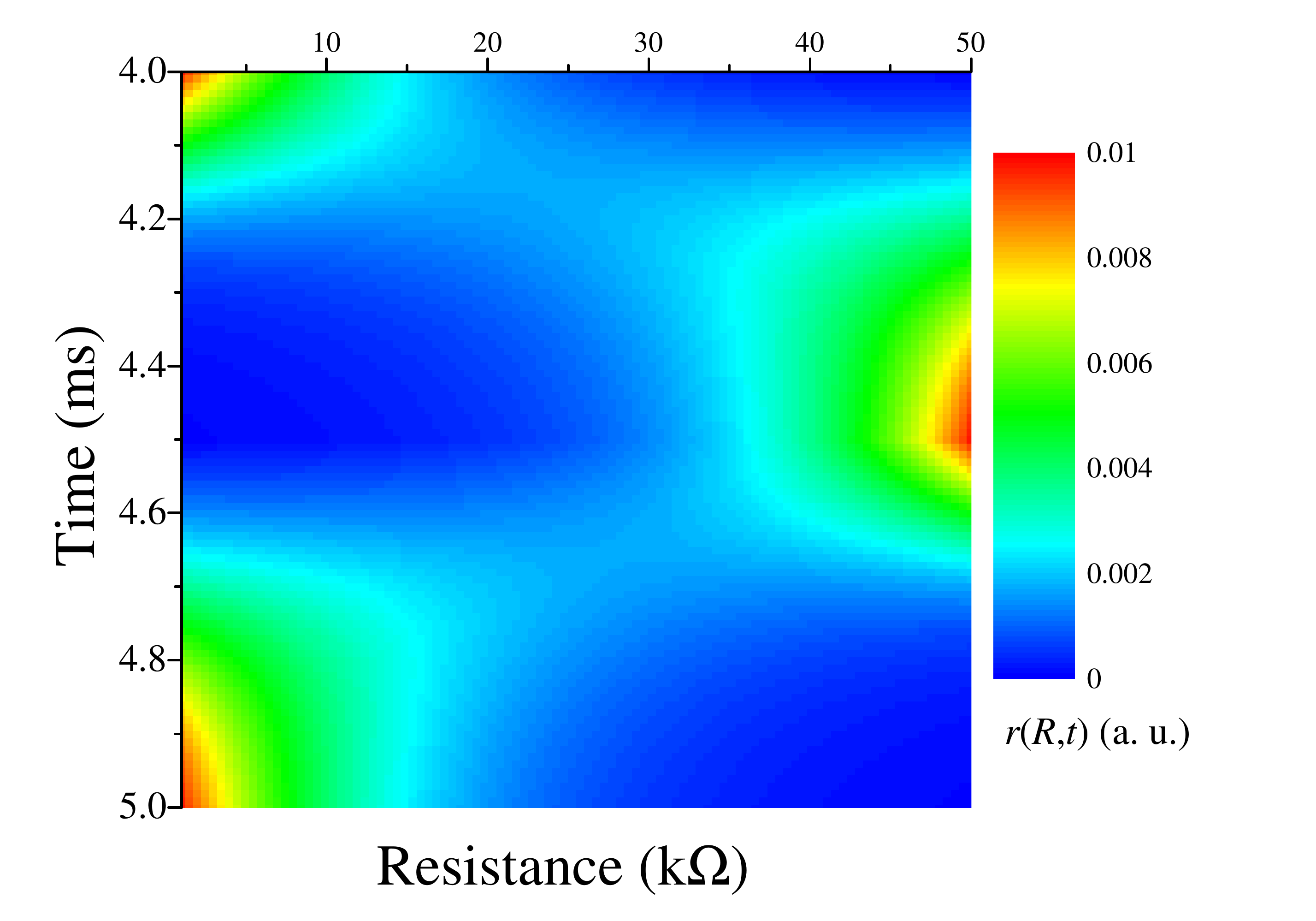}
  \includegraphics[width=0.32\linewidth]{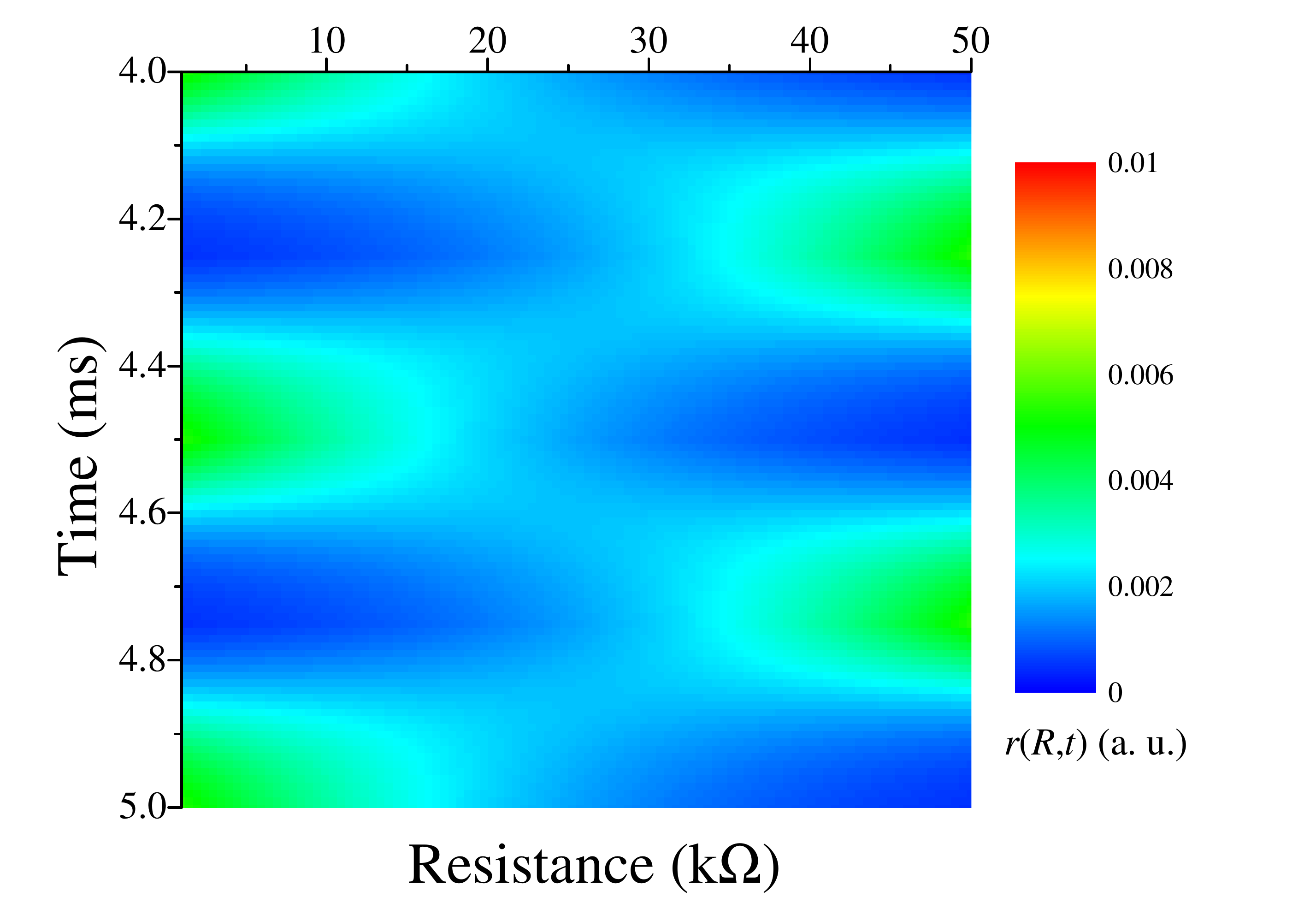}
  \includegraphics[width=0.32\linewidth]{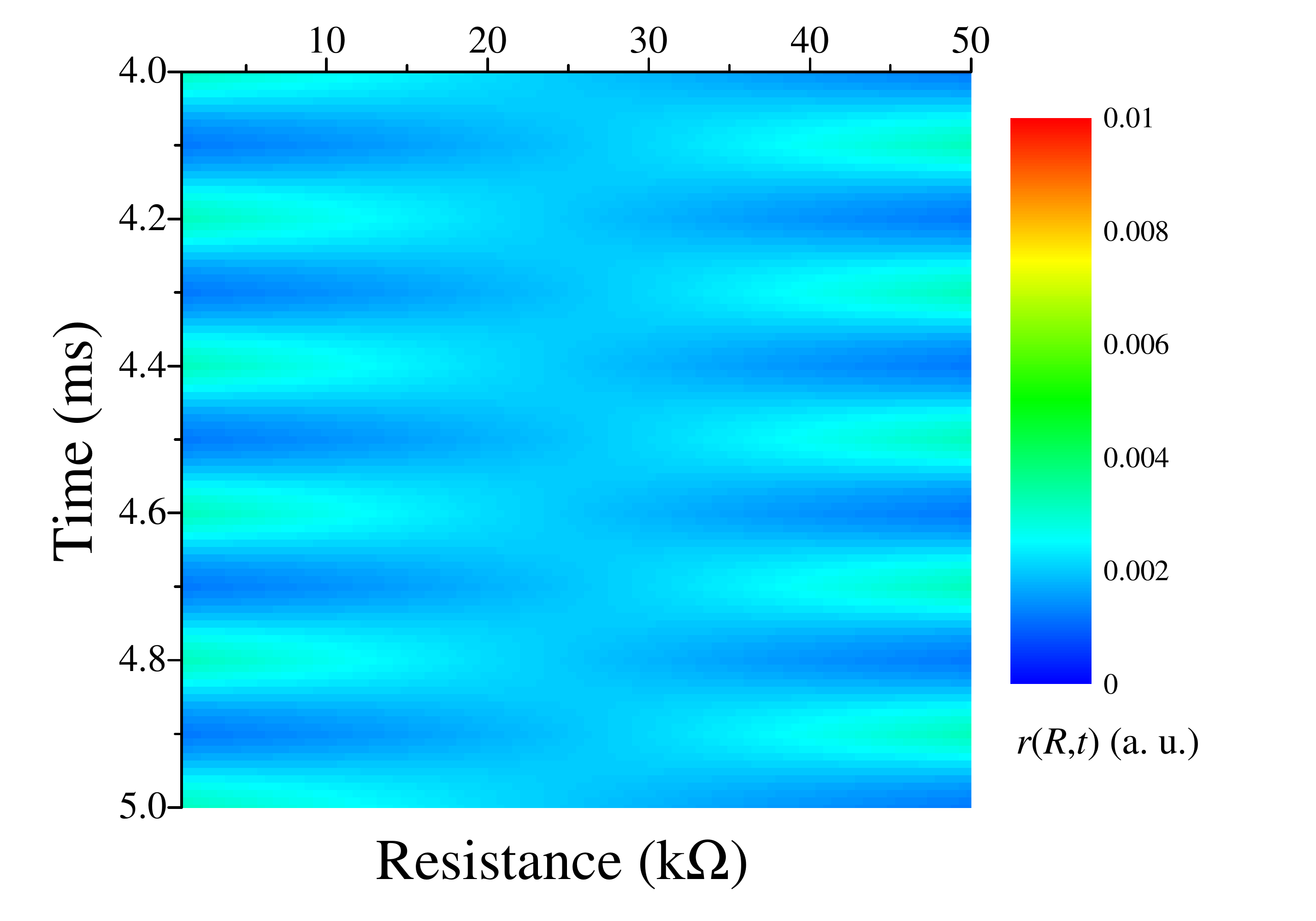} \\
  (d) \hspace{5cm} (e) \hspace{5cm} (f)
  \caption{First model simulations. Response to a sinusoidal voltage $V=V_0 \sin \left( 2\pi f t\right)$: (a) mean current-voltage curves, (b) $\langle R \rangle (t)$ versus voltage, (c) resistance distribution function  at selected times ($f=1$~kHz), and (d)-(f) color plots of $r(R,t)$ at $f=1,2,5$~kHz, respectively. This figure was obtained using the following parameter values: $R_{on}=1$~k$\Omega$, $R_{off}=50$~k$\Omega$, $\alpha_{10}=\alpha_{01}=0.1$~$(\textnormal{s}\cdot \Omega)^{-1}$, and $V_{10}=V_{01}=V_0=1$~V.}
  \label{fig:2}
\end{figure*}

\begin{figure*}
  (a) \includegraphics[width=0.32\linewidth]{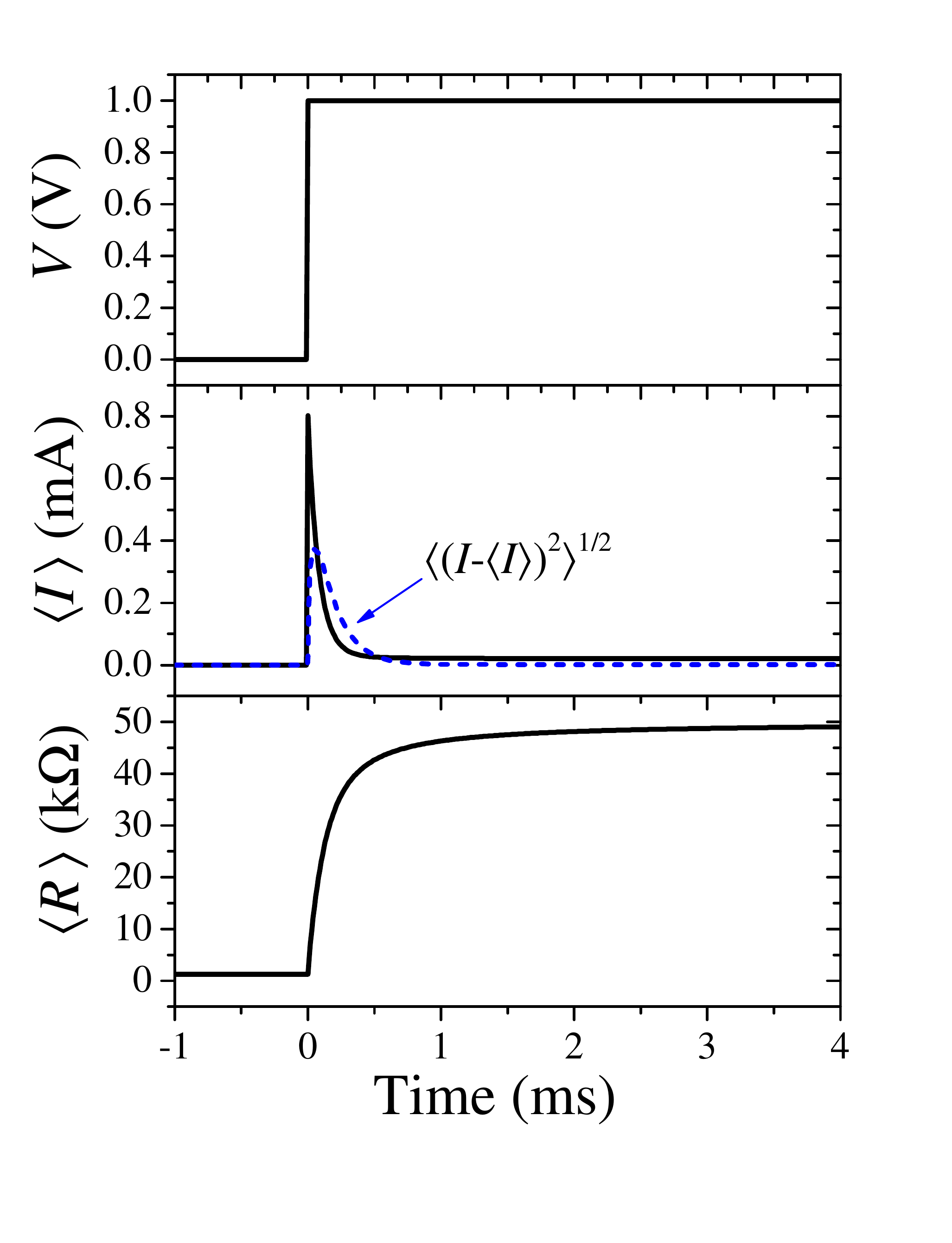}
  (b)\includegraphics[width=0.6\linewidth]{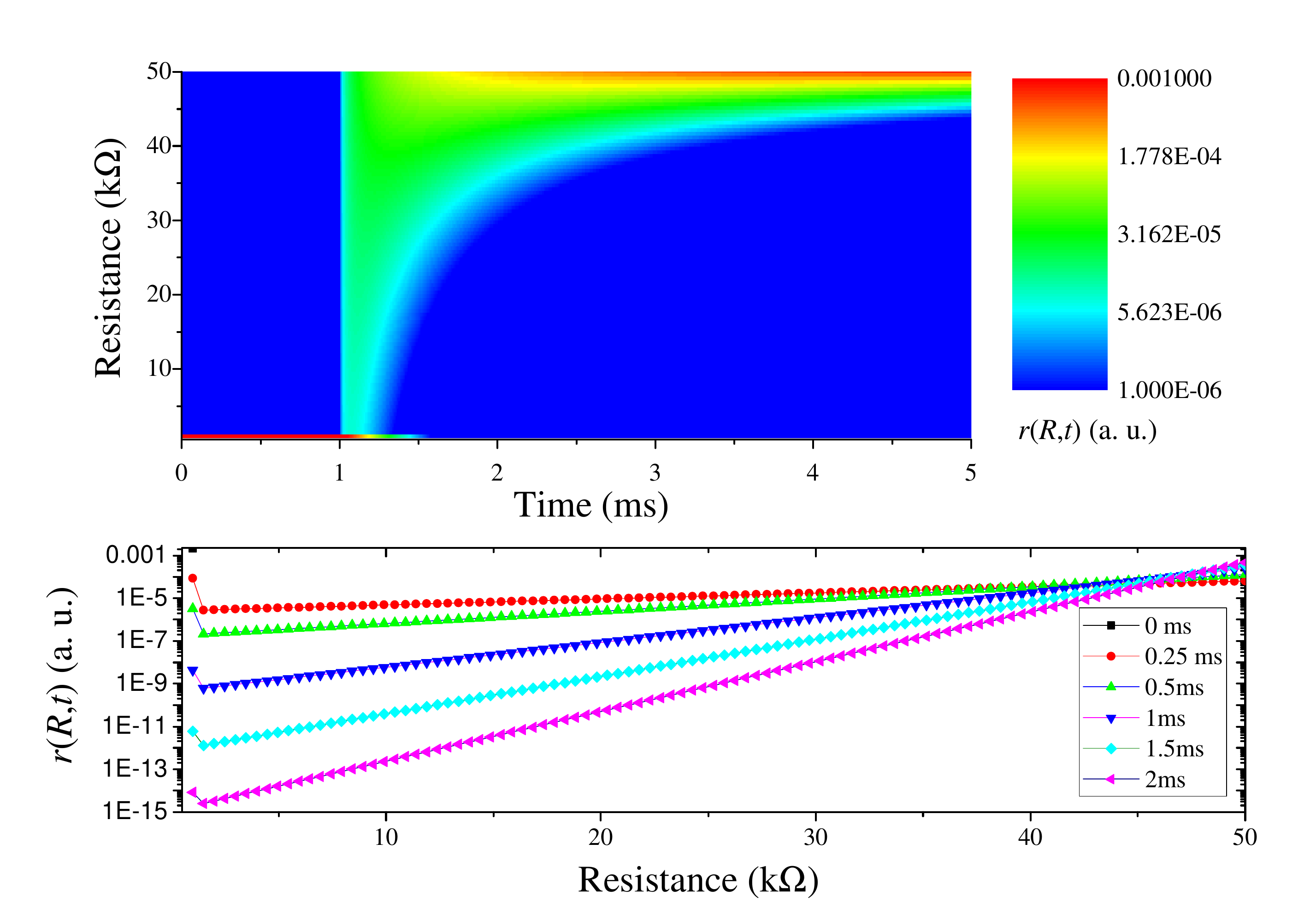}
  \caption{First model simulations. Response to a step-like voltage: (a) Voltage, mean current and mean  resistance as functions of time, (b) color plot (log scale) of $r(R,t)$ (top), and $r(R,t)$ at selected times (bottom). This figure was obtained using the same parameter values as in Fig.~\ref{fig:2} and $r(R,0) =\delta\left( R-R_{on}\right)$.}
  \label{fig:3}
\end{figure*}

To illustrate the above model, we consider two special cases of Eqs.~(\ref{eq:m1}) and (\ref{eq:m2}). In both cases, we will use the resistance $R\in [R_{on},R_{off}]$ as the internal state variable, $x\equiv R$, and the resistance probability distribution function, $r(R,t)$, as the state probability distribution function.
Here, $R_{on}$ and $R_{off}$ denote the ON and OFF state resistance, respectively.

In the first model, we assume a uniform continuous distribution of jumps in the direction defined by the applied voltage. The transition rate density in Eq.~(\ref{eq:m2}) is selected as
 \begin{equation}
 \gamma(R',R,V) = \left\{ \begin{array}{ccl}
\alpha_{10}e^{\frac{|V|}{V_{10}}}&,& V>0, \; R'<R \\
\alpha_{01}e^{\frac{|V|}{V_{01}}}&,& V<0, \; R'>R\;, \\
0 &,& \textnormal{otherwise}
\end{array}\right.
\label{eq:rates}
\end{equation}
where $\alpha_{10}$, $V_{01}$, and  $\alpha_{01}$, $V_{01}$ are the coefficients defining the transition rates in the direction from ON to OFF (10) and OFF to ON (01) resistance state, respectively, and the exponential dependence on voltage is selected based on experimental observations~\cite{jo2009programmable,gaba2013stochastic,naous2021theory}. Below, we use the  notation
$\gamma_{10}=\alpha_{10}\textnormal{exp}(|V|/V_{10})$ corresponding to the top line in the rhs of Eq.~(\ref{eq:rates}).

Fig.~\ref{fig:2} shows results of numerical simulations for the above model in the case of sinusoidal driving. We note that the frequency behavior of current-voltage curves in Fig.~\ref{fig:2}(a) is typical to memristive devices. Within each period, the maximum of the resistance distribution function $r(R,t)$ drifts from $R_{ON}$ to $R_{OFF}$ and back. Note that Fig.~\ref{fig:2}(c) shows the evolution in the
first half-period of periodic driving. Panels~(d)-(f) in Fig.~\ref{fig:2} demonstrate that the resistance distribution function becomes less localized at higher frequencies. Overall, all these results are not unexpected.

Next, we simulate the response to step voltage (Fig.~\ref{fig:3}). According to Fig.~\ref{fig:3}(a), the step voltage switches the resistance from $R_{on}$ to $R_{off}$ in a relatively short time interval. A notable feature is the exponential decay in the resistance probability distribution function as seen in the bottom panel in Fig.~\ref{fig:3}(b).

 The analytical solution of Eq.~(\ref{eq:m2}) with $\gamma$ given by Eq.~(\ref{eq:rates}) and $V=\textnormal{const}$  can be found using the method of  Laplace transform. One can show (see SI) that the general solution for $V >0$  can be written as
\begin{eqnarray}
   r(R,t)=\gamma_{10} t e^{-\gamma_{10}(R_{off}-R)t}
   \int\limits_{R_{on}}^R r(R',0)\textnormal{d}R' \nonumber \\
   +r(R,0)e^{-\gamma_{10}(R_{off}-R)t} \; ,
       \label{eq:solution1}
\end{eqnarray}
where $r(R,0)$ is the initial resistance probability distribution function.

The simple form of the general solution~(\ref{eq:solution1}) allows the detailed analysis of the evolution for any initial resistance probability distribution function $r(R,0)$. % (defined on the interval $[R_{on},R_{off}]$).
Consider, for instance, $r(R,0)=\delta (R-R_{on})$ (the detailed analysis of this case is given in SI).  One can easily see that the probability distribution function decreases purely exponentially at the left edge of the distribution (at $R=R_{on}$) with the highest possible rate $\gamma_{10}(R_{off}-R_{on})$:
\begin{equation}
   r(R_{on},t)=r(R_{on},0)e^{-\gamma_{10}\left(R_{off}-R_{on}\right)t}.
       \label{eq:solution_on}
\end{equation}
At the right edge of the resistance interval, the resistance probability distribution finction growths linearly from $t=0$:
\begin{equation}
   r(R_{off},t)=r(R_{off},0)+\gamma_{10} t.
       \label{eq:solution_off}
\end{equation}
This result corresponds to the fact that at the positive voltage,
the resistance can only increase (or stay constant), see Eq.~(\ref{eq:rates}).  This leads to the accumulation of the probability density at $R=R_{off}$.
It also follows directly from Eq.~(\ref{eq:solution1}) that inside of the interval $[R_{on},R_{off}]$, $r(R,t)$ increases first and then decreases. The decrease is exponenntial with $R$-dependent rate $\gamma_{10}(R_{off}-R)$.
Moreover, the linear growth of probability~(\ref{eq:solution_off}) at  $R_{off}$ implies that the characteristic width of the distribution peak decreases as $(\gamma_{10} t)^{-1}$ asymptotically in time.

Eq.~(\ref{eq:solution1}) allows finding various quantities on average such as the mean resistance
\begin{equation}
   \langle R\rangle (t) =R_{off}-
   \frac{1-e^{-\gamma_{10}(R_{off}-R_{on})t}}{\gamma_{10} t}
       \label{eq:R average uniform1}
\end{equation}
and its variance (see SI).

\begin{figure*}
  (a) \includegraphics[width=0.32\linewidth]{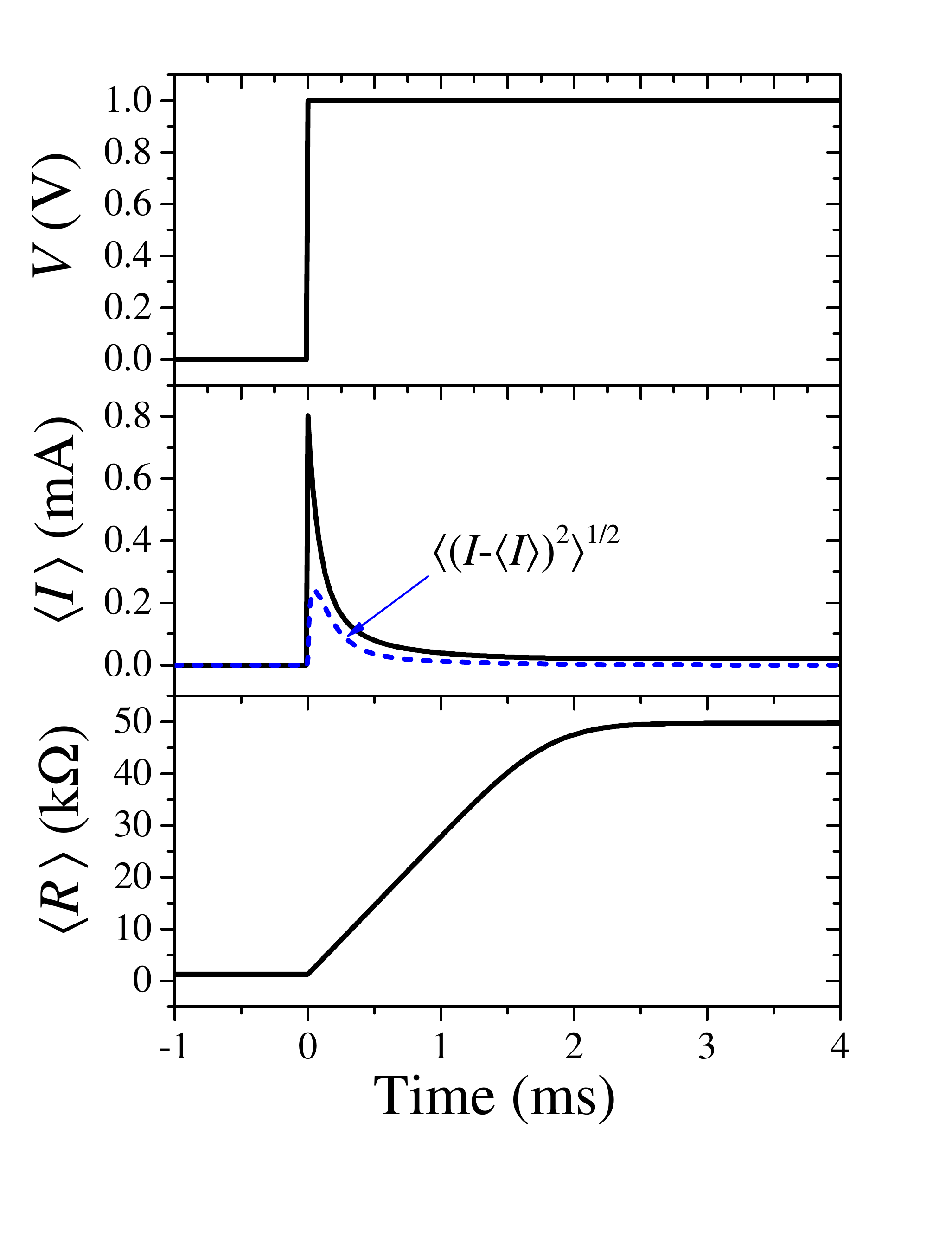}
  (b)\includegraphics[width=0.6\linewidth]{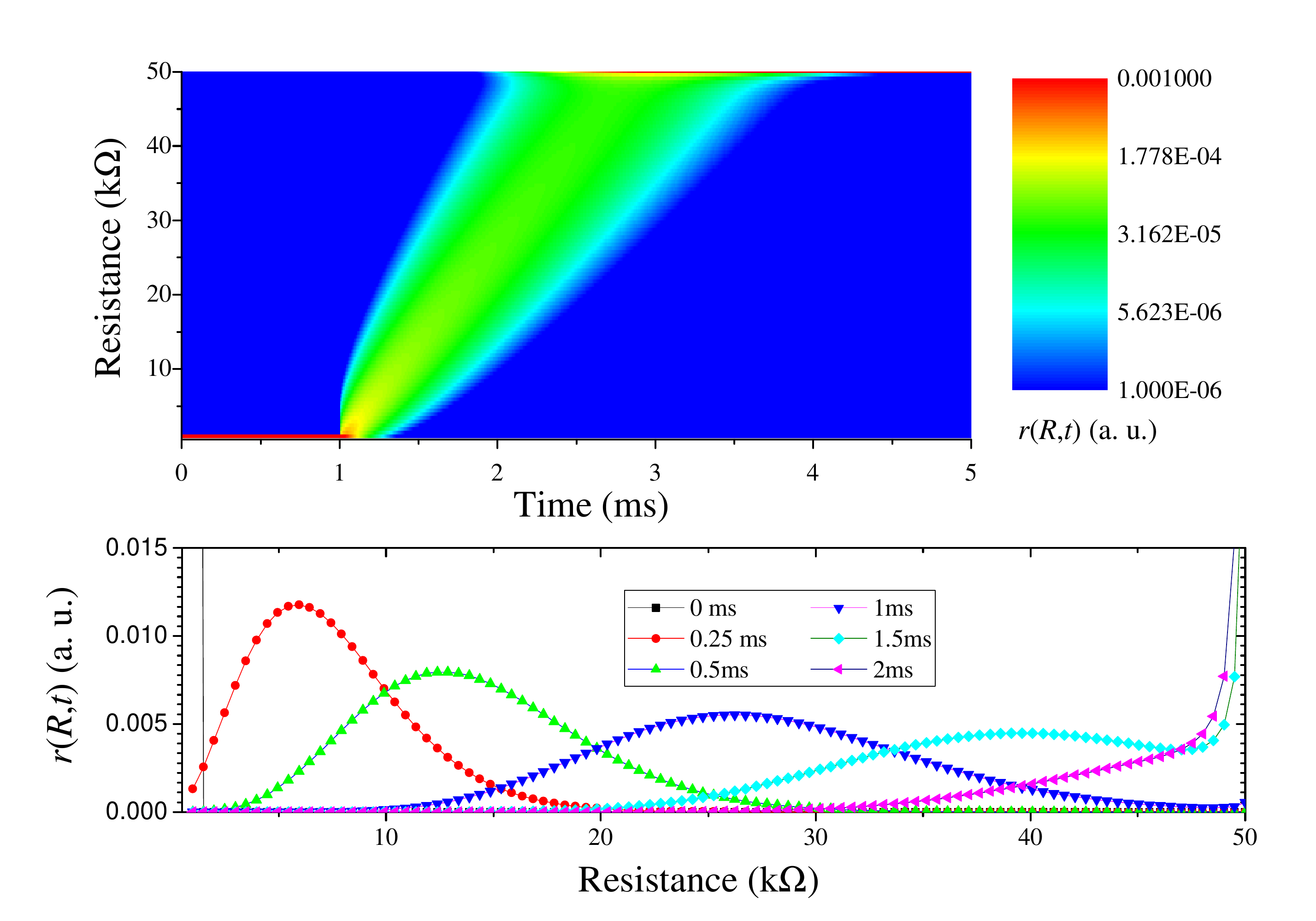}
  \caption{Second model simulations. Response to a step-like voltage: (a) Voltage, mean current and mean resistance as functions of time; (b) color plot (log scale) of $r(R,t)$ (top), and $r(R,t)$ at selected times (bottom). This figure was obtained using the following parameter values: $R_{on}=1$~k$\Omega$, $R_{off}=50$~k$\Omega$, $\alpha_{10}=\alpha_{01}=10$~$(\textnormal{s}\cdot \Omega)^{-1}$, $V_{10}=V_{01}=1$~V, $R_0=1$~k$\Omega$,  and $r(R,0) =\delta\left( R-R_{on}\right)$. }
  \label{fig:4}
\end{figure*}

In the second model we use $R$-dependent transition rates to describe the situation when the shorter jumps are more frequent than longer. The transition rate density is selected as
 \begin{equation}
 \gamma(R,R',V) = \left\{ \begin{array}{ccl}
\alpha_{10}e^{\frac{|V|}{V_{10}}-\frac{|R-R'|}{R_0}}, V>0, \; R<R' \\
\alpha_{01}e^{\frac{|V|}{V_{01}}-\frac{|R-R'|}{R_0}}, V<0, \; R>R'\;, \\
\;\;\;0 \;\; \;\;\;\;\;\;\;\;, \;\;\;\; \textnormal{otherwise}
\end{array}\right.
\label{eq:rates2}
\end{equation}
where $R_0$ defines the average size of the jump. In fact, the first model can be considered as the limiting case of the second model in the limit of $R_0\rightarrow \infty$.

Fig.~\ref{fig:4} shows the response of the second model to step voltage. We highlight a couple of interesting features of this result. First, unlike Fig.~\ref{fig:3}(a), in Fig.~\ref{fig:4}(a) there is a significant interval of the linear growth of resistance (from 0~ms to 1.5~ms). Second,  the resistance probability distribution function is more localized: in the bottom panel of Fig.~\ref{fig:4}(b) there is a clear peak  that shifts from left to right as the resistance switches from $R_{on}$ to $R_{off}$.

Compared to the first model, the analytic analysis of the second one is more involved. Its general solution is complex
(we give it in SI, see Eq.~(\ref{eq:general_solution}) supplemented by Eq.~(\ref{eq:contour_integral})). This exact analytical solution allows the efficient numerical calculation of the resistance probability distribution function $r(R,t)$ at any moment of time for the whole range of parameters, in particular for small values of $R_0$.  Moreover, it allows the derivation of simple asymptotic formulas for interesting limiting cases.

In particular, the solution for $r(R,0)=\delta(R-R_{on})$ (given by  Eq.~(\ref{eq:solution for delta small a}) in SI) allows to calculate directly the mean resistance and its variance  {\it at short times},
\begin{equation}
   \langle R\rangle =R_{on}+R_0^2\gamma_{10} t,
       \label{eq:R average nonuniform}
\end{equation}
and
\begin{equation}
   \langle (R- \langle R\rangle)^2\rangle =
  2R_0^3\gamma_{10} t.
       \label{eq:R variance nonuniform}
\end{equation}
In fact, Eqs.~(\ref{eq:R average nonuniform}), (\ref{eq:R variance nonuniform}) are exact  in the limit of $R_{off}\rightarrow \infty$ in Eq.~(\ref{eq:solution for delta small a}).
Note that Eq.~(\ref{eq:R average nonuniform}) explains
the linear increase of $\langle R \rangle$ in  Fig.~\ref{fig:4}(a).

Additionally, if $\gamma_{10} t (R-R_{on})=\gamma_{10}R_0t\xi\gg 1$, the solution (\ref{eq:solution for delta small a}) can be simplified to
\begin{equation}
  r(R,t)=\frac{(\gamma_{10} R_0t)^{1/4}}{2\sqrt{\pi}\xi^{3/4}R_0}
  e^{-(\sqrt{\xi}-\sqrt{\gamma_{10} R_0t})^2},
       \label{eq:asymptotic}
\end{equation}
where, for shortness, we have introduced  $\xi=(R-R_{on})/R_0$.

From Eq.~(\ref{eq:asymptotic}) we see that far from both boundaries, $R_{on}$ and $R_{off}$,  the resistance probability distribution function $r(R,t)$ tends asymptotically to the bell-shaped distribution (Eq.~(\ref{eq:asymptotic})) at $t\gg 1/(\gamma_{10} R_0\xi)$. The location of the maximum, $\xi_{max}=\xi_{max}(t)$, can be easily calculated from Eq.~(\ref{eq:asymptotic}). At long times, $\gamma_{10} R_0t\gg 1$, we find that $\xi_{max}(t)=\gamma_{10} R_0t-3/2+O(1/t)$. It means that in this regime the distribution  maximum, $\xi_{max}$, propagates at constant velocity $\gamma_{10} R_0^2$ in the $R$-space. The magnitude of the distribution maximum $r(\xi_{max}(t),t)=1/(2R_0\sqrt{\pi\gamma_{10} R_0t})(1+O(1/t))$ decreases as the inverse square root of time. This also means that the characteristic width of the probability distribution increases as the square root of time (to satisfy the normalization condition). Another way to see it is to expand up to quadratic terms with respect to $(\xi-\xi_{max})$ the logarithm  of Eq.~(\ref{eq:asymptotic}). This way we get the following approximated expression for the probability distribution function (\ref{eq:asymptotic}), which is valid in the vicinity of the maximum point $\xi_{max}(t)$ when $\gamma_{10} R_0 t\gg 1$:
 \begin{equation}
  r(R,t)=\frac{1}{2R_0\sqrt{\pi \gamma_{10} R_0t}}
  e^{-\frac{(\xi-\gamma_{10} R_0t+3/2)^2}{4\gamma_{10} R_0t}}.
       \label{eq:Gaussian asymptotic}
\end{equation}
Eq.~(\ref{eq:Gaussian asymptotic}) is the Gaussian distribution with the maximum at  $\xi_{max}\approx\gamma_{10} R_0 t -3/2$ and standard deviation $\sqrt{2\gamma_{10} R_0 t}$.

Thus we see that the evolution of the initial delta function distribution $r(R,t)=\delta(R-R_{on})$ involves three stages (when $|R_{off}-R_{on}|\gg R_0$).
At the first stage, $0<t\lesssim (\gamma_{10}R_0)^{-1}$, the ``running wave'', which is described by the second term in Eq.~(\ref{eq:solution for delta small a}), is formed near $R_{on}$. At the second stage, $(\gamma_{10}R_0)^{-1} \lesssim t \lesssim (R_{off}-R_{on})/(\gamma_{10} R_0^2)$, as the first term in Eq.~(\ref{eq:solution for delta small a}) becomes exponentially small, the second term, which can be approximated by Eq.~(\ref{eq:asymptotic}), describes the propagation of the ``wave'' at  constant velocity $\gamma_{10} R_0^2$ and its broadening as $2R_0\sqrt{\pi \gamma_{10}R_0t}$ in $R$-space. At the beginning of the last third stage of evolution, $t\gtrsim (R_{off}-R_{on})/(\gamma_{10} R_0^2)$, the ``wave'' reaches $R_{off}$. At this moment of time, the probability distribution has a characteristic width of $2\sqrt{\pi R_0(R_{off}-R_{on})}$, which is much larger than $R_0$ (by a factor of $\sim 2\sqrt{\pi(R_{off}-R_{on})/R_0}\gg 1$), and, at the same time, much smaller than $(R_{off}-R_{on})$.Therefore, the probability distribution stays localized the whole evolution time. Eventually, the magnitude of the probability distribution  at $R_{off}$ starts to grow linearly in time, $r(R_{off},t)\sim\gamma_{10} t$, as it follows from the asymptotic behavior of  Laplace transform $ \tilde r(R_{off},p)\sim\gamma_{10}/p^2$ when $p\rightarrow 0$. At the same time, the probability distribution approaches the delta function $\delta(R-R_{off})$.

In conclusion, we have developed a fundamentally different probabilistic model of memristive devices that takes into account the cycle-to-cycle variability in their response. Our model differs conceptually from the conventional memristive models in its statistical approach to the description of resistance switching phenomenon. Overall, this work provides a new tool for the analysis of stochastic memristive devices and their circuits.

\begin{acknowledgement}

The authors are thankful to each other for productive collaboration resulted in this paper.

\end{acknowledgement}

\begin{suppinfo}

%This will usually read something like: ``Experimental procedures and
%characterization data for all new compounds. The class will
%automatically add a sentence pointing to the information on-line:

Details of the solutions; extension of the model to several internal state variables.

\end{suppinfo}

%\bibliography{references}

\providecommand{\latin}[1]{#1}
\makeatletter
\providecommand{\doi}
  {\begingroup\let\do\@makeother\dospecials
  \catcode`\{=1 \catcode`\}=2 \doi@aux}
\providecommand{\doi@aux}[1]{\endgroup\texttt{#1}}
\makeatother
\providecommand*\mcitethebibliography{\thebibliography}
\csname @ifundefined\endcsname{endmcitethebibliography}
  {\let\endmcitethebibliography\endthebibliography}{}

\clearpage
\newpage

\onecolumn

\renewcommand{\theequation}{S.\arabic{equation}}
\setcounter{equation}{0}
\setcounter{page}{1}

\begin{center}
    {\LARGE Supporting Information}
\end{center}

\vspace{5mm}

\noindent {\LARGE \bf A probabilistic model of resistance jumps in
memristive devices}

\vspace{5mm}

\noindent Valeriy A. Slipko$^1$ and Yuriy V. Pershin$^2$

\vspace{5mm}

\noindent $^1${\it Institute of Physics, Opole University, Opole 45-052, Poland}

\noindent $^2${\it Department of Physics and Astronomy, University of South Carolina, Columbia, SC 29208, USA}

\vspace{5mm}

\noindent E-mails: vslipko@uni.opole.pl; pershin@physics.sc.edu

\section{Uniform distribution of jumps}

\subsection{Laplace transform solution}
To solve the integro-differential equation~(\ref{eq:m2}) with $\gamma$ given by Eq.~(\ref{eq:rates}) at constant $V>0$, we introduce the Laplace transform in the time domain as $\tilde r(R,p)=\int_0^\infty r(R,t)\exp(-pt)\textnormal{d}t$. Applying the Laplace transformation to Eq.~(\ref{eq:m2}) one obtains
\begin{equation}
    \left[p+\gamma_{10} (R_{off}-R)\right]\tilde r(R,p)=
    \gamma_{10}\int\limits_{R_{on}}^R \tilde r(R',p)\textnormal{d}R'+r(R,0),
    \label{eq:LT_integral}
\end{equation}
where $r(R,0)\equiv r(R, t=0)$ is the initial resistance probability distribution function.

By differentiating Eq.~(\ref{eq:LT_integral}) with respect to $R$, we arrive at a first-order non-homogeneous linear differential equation with the following general solution
\begin{equation}
   \tilde r(R,p)=\frac{r(R,0)}{\left[p+\gamma_{10} (R_{off}-R)\right]}+
  \frac{C(p)+\gamma_{10}\int\limits_{R_{on}}^R r(R',0)\textnormal{d}R'}{\left[p+\gamma_{10} (R_{off}-R)\right]^2},
    \label{eq:LT_solution}
\end{equation}
where $C(p)$ is an arbitrary function.
Consider now Eq.~(\ref{eq:LT_integral}) at $R=R_{on}$. Its solution is $ \tilde r(R_{on},p)=r(R_{on},0)/\left[p+\gamma_{10} (R_{off}-R_{on})\right]$. Comparing this expression with Eq.~(\ref{eq:LT_solution}) at $R=R_{on}$, we find that $C(p)=0$. The inverse Laplace transform leads to
the following general solution of Eq.~(\ref{eq:m2}):
\begin{equation}
   r(R,t)=\gamma_{10} t e^{-\gamma_{10}(R_{off}-R)t}
   \int\limits_{R_{on}}^R r(R',0)\textnormal{d}R'
   +r(R,0)e^{-\gamma_{10}(R_{off}-R)t}  \;\; .
       \label{eq:solution1SI}
\end{equation}

\subsection{Solution analysis}

To illustrate the solution~(\ref{eq:solution1SI}), let us consider the evolution of the resistance initially localized at $R=R_{on}$. Such situation is described by the initial resistance probability distribution function $r(R,0)=\delta(R-R_{on})$, where $\delta$ is the Dirac delta function.  From Eq.~(\ref{eq:solution1SI}) it follows that at any moment of time the probability distribution is the sum of two terms: \begin{equation}
   r(R,t)=
   \delta(R-R_{on})e^{-\gamma_{10}(R_{off}-R_{on})t}+\gamma_{10} t e^{-\gamma_{10}(R_{off}-R)t}.
       \label{eq:delta solution1}
\end{equation}
The first term corresponds to the initial distribution function, whose magnitude decreases exponentially in time with the highest possible rate $\gamma_{10}(R_{off}-R_{on})$ in the system (due to transitions into the continuum of states $R_{on}<R\leq R_{off}$). The second term grows at the initial period of time until $t(R)=(\gamma_{10}(R_{off}-R))^{-1}$. Later, at $t>t(R)$, the second term  asymptotically decreases to zero everywhere except of $R=R_{off}$, where the probability distribution function accumulates with time.

These properties of the distribution function are reflected in the behavior of the mean resistance as a function of time that can be calculated by using Eq.~(\ref{eq:delta solution1}):
\begin{equation}
   \langle R\rangle (t) =\int\limits_{R_{on}}^{R_{off}} r(R,t)R\textnormal{d}R=R_{off}-
   \frac{1-e^{-\gamma_{10}(R_{off}-R_{on})t}}{\gamma_{10} t}.
       \label{eq:R average uniform}
\end{equation}
The mean resistance changes linearly with time at short times $\gamma_{10}(R_{off}-R_{on})t\ll 1$ starting at the value $R_{on}$
\begin{equation}
   \langle R\rangle (t) =R_{on}+\frac{\gamma_{10} t}{2}(R_{off}-R_{on})^2,~t\rightarrow{0},
       \label{eq:R average uniform small time}
\end{equation}
and reaching asymptotically the value $R_{off}$ at long times $\gamma_{10}(R_{off}-R_{on})t\gg 1$ in accordance with the law
\begin{equation}
   \langle R\rangle (t) =R_{off}-\frac{1}{\gamma_{10} t},~t\rightarrow{\infty}.
       \label{eq:R average uniform large time}
\end{equation}

Similarly we can calculate the variance of resistance
\begin{equation}
   \langle (R- \langle R\rangle)^2\rangle =
   \frac{
   1-
   2\gamma_{10} t (R_{off}-R_{on})e^{-\gamma_{10}(R_{off}-R_{on})t}-
    e^{-2\gamma_{10}(R_{off}-R_{on})t}
   }{(\gamma_{10} t)^2},
       \label{eq:R variance uniform}
\end{equation}
with the following asymptotic behavior at short and long times
\begin{equation}
  \langle (R- \langle R\rangle)^2\rangle =
   \frac{\gamma_{10} t}{3}(R_{off}-R_{on})^3,~\text{when}~\gamma_{10}(R_{off}-R_{on})t\ll 1,
       \label{eq:R variance uniform small time}
\end{equation}
and
\begin{equation}
   \langle (R- \langle R\rangle)^2\rangle =\frac{1}{(\gamma_{10} t)^2},
   ~\text{when}~\gamma_{10}(R_{off}-R_{on})t\gg 1.
       \label{eq:R variance uniform large time}
\end{equation}

Thus we see that being initially localized at $R_{on}$, the resistance probability distribution expands into $R$-space due to transitions into all states with $R>R_{on}$.  The distribution goes through the transient period of time in the region $(R_{on}, R_{off})$, and it gradually accumulates in the vicinity of the boundary point $R_{off}$, asymptotically in time.

It is interesting to note that precise analytical solution (\ref{eq:solution1SI}) can be used to express
the mean resistance at any moment of time $t$ through the resistance probability distribution function at $t=0$:
\begin{equation}
   \langle R\rangle (t) =R_{off}-\frac{1}{\gamma_{10}t}+\frac{1}{\gamma_{10}t}
   \int\limits_{R_{on}}^{R_{off}} \textnormal{d}R~r(R,0)
   e^{-\gamma_{10}(R_{off}-R)t}.
       \label{eq:R average uniform general}
\end{equation}
By expending the exponential function $e^{\gamma_{10}Rt}$ in Eq.~(\ref{eq:R average uniform general}), we can demonstrate how the mean resistance $\langle R\rangle (t)$ explicitly depends on all the moments
$\langle R\rangle (0)$, $\langle R^2\rangle (0)$, ... at the initial moment of time $t=0$:
\begin{equation}
   \langle R\rangle (t) =R_{off}+\frac{e^{-\gamma_{10}R_{off}t}-1}{\gamma_{10}t}+
   e^{-\gamma_{10}R_{off}t}
   \sum\limits_{n=1}^{+\infty} \frac{(\gamma_{10}t)^{n-1}}{n!}\langle R^n\rangle (0).
       \label{eq:R average uniform general moments}
\end{equation}
Because all the moments can be independently specified, the mean resistance depends  in principle on the infinite number of parameters (the initial moments). Consequently, in general, there is no differential equation of finite order, which could describe the evolution of the mean resistance for an arbitrary initial condition. But if we limit the class of the initial conditions considering, for example, only deterministic initial conditions, $r(R,0)=\delta(R-R_i)$ with some initial resistance $R_i$, then the corresponding dynamic differential equations for the mean resistance can be found.

\section{Non-uniform distribution of jumps}

\subsection{Laplace transform solution}

In this section we consider the dynamics of the second model in the presence of a positive constant voltage.
The substitution of the transition rate $\gamma(R',R,V_M)$ from Eq.~(\ref{eq:rates2}) into Eq.~(\ref{eq:m2})  leads to
\begin{equation}
    \frac{\partial g(R,t)}{\partial t}=\gamma_{10}\int\limits_{R_{on}}^R g(R',t)\textnormal{d}R'-\Gamma(R)g(R,t),
    \label{eq:master_g}
\end{equation}
where $g(R,t)=r(R,t)e^{R/R_0}$ is a new (non-normalized) resistance probability distribution function. In the above equation,
we have also introduced  the rate
\begin{equation}
    \Gamma(R)=\gamma_{10} R_0\left(1-\exp\left\{(R-R_{off})/R_0\right\}\right).
    \label{eq:rates3}
\end{equation}

In the Laplace domain, the general solution of Eq.~(\ref{eq:master_g}) with arbitrary initial condition $r(R,0)$ is given by
\begin{equation}
   \tilde r(R,p)=\frac{r(R,0)}{p+\Gamma (R)}+
   \int\limits_{R_{on}}^R \Phi(p,R,R_1) r(R_1,0)\textnormal{d}R_1,
   \label{eq:LT_solution2}
\end{equation}
where $\Phi(p,R,R_1)$ is the regular single-valued on the whole complex plane, except for the branch cut $[-\Gamma(R_1),-\Gamma(R)]$, branch of the analytic function of complex variable $p$:
\begin{equation}
    \Phi(p,R,R_1)=\frac{\gamma_{10}}{(p+\Gamma(R))^2}
    \exp\left\{
    \frac{p}{p+R_0\gamma_{10}}
    \left[
    \ln \frac{p+\Gamma(R)}{p+\Gamma(R_1)}-\frac{R-R_1}{R_0}
    \right]
    \right\},
    \label{eq:LT_F}
\end{equation}
defined by the condition that the logarithm function is real, $\Im\left(\ln \frac{p+\Gamma(R)}{p+\Gamma(R_1)}\right)$=0, when $p>0$. It can be easily verified that such a branch has a removable singularity at $p=-R_0\gamma_{10}$. Moreover, the function $\Phi(p,R,R_1)$ is analytic at infinity with the following asymptotic behavior: $\Phi(p,R,R_1)=\gamma_{10} \textnormal{exp}\left((R_1-R)/R_0\right)/p^2$ as $p\rightarrow\infty$.

By using these properties of the Laplace transform, we can present the inverse Laplace transform as the integral in the complex $p$-plane along any curve $C$, which encircles the branch cut in the positive (counterclockwise) direction. For further calculations, it is convenient to make the linear fractional transformation $\zeta=R_0\gamma_{10}/(p+\Gamma(R))$.
As a result of this transformation,  the general solution of Eq.~(\ref{eq:master_g}) can be presented as
\begin{equation}
   r(R,t)=
    r(R,0)e^{-\Gamma(R)t}+
    e^{-\Gamma(R)t}
   \int\limits_{R_{on}}^R \frac{\textnormal{d}R_1}{R_0}r(R_1,0)
   \exp{\left\{-\frac{R-R_1}{R_0}\right\}}
   I(t,R,R_1),
       \label{eq:general_solution}
\end{equation}
where we have introduced the contour integral over the unit circle in the counter-clockwise direction
\begin{equation}
   I(t,R,R_1)=\frac{1}{2\pi i}
   \oint\limits_{|\zeta|=1} \frac{\textnormal{d}\zeta}{1+[\delta-\delta_1]\zeta}
   \exp{\left\{
   \frac{\gamma_{10} R_0 t}{\zeta}+
   \frac{\zeta\left(\ln{[1+(\delta-\delta_1)\zeta]}+\frac{R-R_1}{R_0} \right)}{1+\zeta\delta}
   \right\}},
       \label{eq:contour_integral}
\end{equation}
where $\delta=\delta(R)=\textnormal{exp}\left((R-R_{off})/R_0\right)$, and $\delta_1=\delta(R_1)$.

While analyzing the integrand in Eq.~(\ref{eq:contour_integral}), it is useful to note that $0<\delta_1<\delta\leq 1$ since $ R_{on}\leq R_1<R\leq R_{off}$.

\subsection{Solution analysis}

First of all, let us check that solution given by Eqs.~(\ref{eq:general_solution}) and (\ref{eq:contour_integral}) coincides with solution~(\ref{eq:solution1}) in the case when $(R_{off}-R_{on})\ll R_0$. For this purpose we have to replace $\delta$ and $\delta_1$ with $1$ in Eq.~(\ref{eq:contour_integral}). As a result, we find that
$$I(t,R,R_1)=\frac{1}{2\pi i}
   \oint\limits_{|\zeta|=1}
   \textnormal{d}\zeta
     \exp{\left\{
   \frac{\gamma_{10} R_0 t}{\zeta}
   \right\}}=\gamma_{10} R_0 t, ~R_0\rightarrow\infty.$$
   This result along with the observation that $\Gamma(R)\rightarrow\gamma_{10}(R_{off}-R)$, when $R_0\rightarrow\infty$, shows that general solution~(\ref{eq:general_solution}), (\ref{eq:contour_integral}) transforms to
   solution~(\ref{eq:solution1}) in the limiting case $(R_{off}-R_{on})\ll R_0$ as it should be.

In the opposite limiting case $(R_{off}-R_{on})\gg R_0$, the general solution~(\ref{eq:general_solution}), (\ref{eq:contour_integral}) also can be simplified in the domain far from the right end $R_{off}$, where
$|R_{off}-R|\gg R_0$.
In this case both $\delta$ and $\delta_1$ are exponentially small, so we can replace them with zeroes  in
Eq.~(\ref{eq:contour_integral}). Thus we obtain the following approximate expression
\begin{equation}
I(t,R,R_1)=\frac{1}{2\pi i}
   \oint\limits_{|\zeta|=1}
   \textnormal{d}\zeta
     \exp{\left\{
   \frac{\gamma_{10} R_0 t}{\zeta}+\zeta\frac{(R-R_1)}{R_0}
   \right\}}= R_0\gamma_{10} t\sum_{k=0}^{+\infty}
   \frac{[\gamma_{10} t(R-R_1)]^k}{k!(k+1)!},
    \label{eq:solution small a}
\end{equation}
when $R_0\rightarrow 0$.
By using the power series expansion of the modified Bessel function of the first kind,
$\sum_{k=0}^{+\infty}
   \frac{x^k}{k!(k+1)!}=I_1(2\sqrt{x})/\sqrt{x}$,
we get
\begin{equation}
I(t,R,R_1)=R_0 \gamma_{10} t \frac{I_1(2\sqrt{\gamma_{10} t(R-R_1)})}{\sqrt{\gamma_{10} t(R-R_1)}},
\textnormal{when } |R_{off}-R|\gg R_0.
    \label{eq:solution small a 2}
\end{equation}

Eq.~(\ref{eq:solution small a 2}) along with Eq.~(\ref{eq:general_solution}) allow us to write
the general solution of Eq.~(\ref{eq:m2}) in the following compact form
\begin{equation}
   r(R,t)=
    r(R,0)e^{-\gamma_{10} R_0t}+
   \gamma_{10} t e^{-\gamma_{10} R_0t}
   \int\limits_{R_{on}}^R \textnormal{d}R_1 r(R_1,0)
   e^{-\frac{R-R_1}{R_0}}
   \frac{I_1(2\sqrt{\gamma_{10} t(R-R_1)})}{\sqrt{\gamma_{10} t(R-R_1)}},
       \label{eq:general solution small a}
\end{equation}
for the domain where $|R_{off}-R|\gg R_0$.

To illustrate the evolution of the distribution function determined by solution~(\ref{eq:general solution small a}), we consider the initial condition in the form of the delta function $r(R,0)=\delta (R-R_{on})$.
The delta function initial distribution corresponds to the initial state of memristor when resistivity $R$ equals $R_{on}$ with probability 1. In this case, we immediately get from Eq.~(\ref{eq:general solution small a})
\begin{equation}
   r(R,t)=
  e^{-\gamma_{10} R_0t}  \delta(R-R_{on})+
   \gamma_{10} t e^{-\gamma_{10} R_0t
   -\frac{R-R_{on}}{R_0}}
   \frac{I_1(2\sqrt{\gamma_{10} t(R-R_{on})})}{\sqrt{\gamma_{10} t(R-R_{on})}},
       \label{eq:solution for delta small a}
\end{equation}
for the domain where $ |R_{off}-R|\gg R_0$.

\section{Several internal state variables}

Here we briefly discuss the extension of Eqs.~(\ref{eq:m1})-(\ref{eq:m2}) model to devices described by $N>1$ internal state variables $\{x_1,...,x_N\}\equiv {\bf x}$. Here, the bold font is used to denote a vector.

In the case of several internal state variables, the state probability distribution function is $p({\bf x},t)$, and its normalization is given by
\begin{equation}
    \int\limits_{a_1}^{b_1}...\int\limits_{a_N}^{b_N} p({\bf x},t)\textnormal{d}x_1...\textnormal{d}x_N=1.
\label{eq:Nnorm}
\end{equation}
The limits of integration in the above integrals depend on the physical nature of state variables and are generally different for different variables.

The first equation in the model is exactly the same as Eq.~(\ref{eq:m1}). The only difference is that the mean current is evaluated with the help of $N$ integrals
 \begin{equation}
   \left<I \right>(t)= \int\limits_{a_1}^{b_1}...\int\limits_{a_N}^{b_N} \frac{V(t)}{R({\bf x},t)} p({\bf x},t)\textnormal{d}x_1...\textnormal{d}x_N.
\label{eq:Imean}
\end{equation}

The evolution of $p({\bf x},t)$ is represented by
\begin{eqnarray}
    \frac{\partial p({\bf x},t)}{\partial t}=\int\limits_{a_1}^{b_2}...\int\limits_{a_N}^{b_N}\left[\gamma({\bf x}',{\bf x},V({\bf x}'))p({\bf x}',t) -p({\bf x},t)\gamma({\bf x},{\bf x}',V({\bf x}
    ))\right]\textnormal{d}x'_1...\textnormal{d}x'_N,
    \label{eq:m2N}
\end{eqnarray}
where $\gamma({\bf y},{\bf z},V({\bf y}))$ is the voltage-dependent transition rate density from ${\bf y}$ to ${\bf z}$, and $V({\bf y})$ is the voltage across the device in state ${\bf y}$.

In the situations, when the probability of the simultaneous change of two or more state variables in a single jump is infinitesimal, the transition rate may be represented by the sum of terms like $\gamma(y_1,z_1,V({\bf y}))\delta(y_2-z_2)...\delta(y_N-z_N)$, etc.

\clearpage
\setcounter{page}{1}

\end{document}